\documentstyle[aps,preprint]{revtex}
\begin{document}
\title{{\bf Quantum correction to
thermodynamical entropy of black hole}}
\author{A. Ghosh and P. Mitra
\footnote{e-mail addresses amit, mitra@tnp.saha.ernet.in}}
\address{Saha Institute of Nuclear Physics\\
Block AF, Bidhannagar\\
Calcutta 700 064, INDIA}
\date{gr-qc/9706054}
\maketitle

\begin{abstract}{The  entropy  of  a  black hole can differ from a
quarter  of  the  area of the horizon because of quantum corrections.
The  correction  is   related   to   the
contribution  to  the  Euclidean functional integral from quantum
fluctuations but is not simply equal to the correction to the effective
action. A  (2+1)
dimensional rotating black hole is explicitly considered.}
\end{abstract}

\section{Introduction}
It  has long been known that classical black hole physics has a set of laws
parallel to  the  laws  of  thermodynamics \cite{BCH}.  By  virtue  of  this
parallelism,  the  area  of  the horizon of a black hole was
interpreted as its entropy \cite{Bek}.  After  the  discovery  of
Hawking  radiation  and  the  development of a semiclassical concept of
temperature for black holes, this analogy  became more well-defined
and  at  present a quarter of the area of the horizon is supposed
to be a quantitative measure of the entropy at the semiclassical level.
While  this  refers  to  the  thermodynamic entropy, it should be
mentioned that recently a statistical interpretation of
this  expression in terms of a counting of microscopic states has  been
suggested \cite{strom}.

As mentioned above, the standard expressions for the  temperature
and entropy are semiclassical ones. It would be interesting to see
what deviations from the  standard
expression   for   the   entropy arise because of higher quantum corrections.
This will be useful for comparison with statistical
computations when they get more refined.

For the  simplest  black  hole,  namely  the  one  discovered  by
Sch\-warzschild, the Hawking temperature is given by
\begin{eqnarray}
T={\hbar\over 8\pi GM},
\end{eqnarray}
where  $M$  is the mass of the black hole. Accordingly, the first
law of {\it thermodynamics} can be written as
\begin{eqnarray}
dM=TdS={\hbar\over 8\pi GM}dS,
\end{eqnarray}
which shows that the entropy must be $4\pi GM^2/\hbar$ upto an  additive
constant,  {\it  i.e.},  essentially  a  quarter  of the area.
While this  result  is  obtained
directly  from  thermodynamics, the expression used
for the temperature in the  above argument is a semiclassical one
and   therefore subject to further quantum  corrections. As such,
the expression for the entropy too must be approximate. This  can
be seen  more  clearly  in  the  Euclidean  approach  to  entropy
\cite{GH},  where  the  functional  integral  is  semiclassically
evaluated. The entropy is given essentially  by  the
effective action, which can be approximated by the on-shell classical action,
equal to a quarter of the area. This action
naturally acquires quantum corrections.

In this paper we  shall consider
the correction term. The correction  to  the
entropy  can be obtained directly
by   calculating   the   functional    integral,  {\it i.e.,} the
effective action, and interpreting it in terms of the thermodynamic
potential related to a grand canonical ensemble,
but the value is not equal to the correction to
the effective action although in the literature the words entropy
and action have often been used synonymously.

As the Schwarzschild black hole does not involve any
charge or angular momentum, it is comparatively
simple. It is more instructive to
consider a rotating  black hole
in (2+1) dimensions \cite{BTZ}. In this case a
quantum  correction  to  the  functional integral is
known  \cite{CT},  so  that one can determine
the corrected entropy. As mentioned above,
it turns out to be  different  from
what one would imagine on the basis of the usual relation between
the action and the entropy.

\section{Rotating black hole in 2+1 dimensions}

A 2+1 dimensional black hole has been studied in \cite{BTZ}. The
rotating version is described by the metric
\begin{eqnarray}
ds^2=-f^2dt^2 +f^{-2}dr^2+r^2(d\phi-{4GJ\over r^2}dt)^2,
\end{eqnarray}
where
\begin{eqnarray}
f^2=-8GM+{r^2\over l^2}+{16G^2J^2\over r^2},
\end{eqnarray}
and $l^{-2}$ is a cosmological constant
\cite{BTZ}. The outer horizon is at $r_+$ where
\begin{eqnarray}
r_+^2=4GMl^2\left[1+\sqrt{1-{J^2\over M^2l^2}}\right].\label{r}
\end{eqnarray}
The Hawking temperature is
\begin{eqnarray}
T={\hbar(r_+^2-4GMl^2)\over \pi r_+l^2}.
\end{eqnarray}
The first law of black hole physics can be written in the
form
\begin{eqnarray}
Td({\pi r_+\over 2\hbar G})=dM-\Omega dJ.\label{A2}
\end{eqnarray}
The {\it area} is $2\pi r_+$ because of the 1-dimensional nature of
the horizon here. This
equation can be directly checked from the expression for $r_+$.
The analogue of the chemical potential for the angular momentum is
\begin{eqnarray}
\Omega=\left.{\partial M\over\partial J}\right|_A={4GJ\over r_+^2}.
\end{eqnarray}
Comparing (\ref{A2}) with the first law of thermodynamics,
we can write
\begin{eqnarray}
S= ({\pi r_+\over 2\hbar G}).
\end{eqnarray}
As mentioned before,  this  can  also  be  understood  from  a functional
integral approach. At
the semiclassical level, the on-shell euclidean action is given by
\begin{eqnarray}
-{I_E^{0}\over\hbar}&=&\int d\tau({r_+^2\over 4\hbar Gl^2}-{M\over
\hbar})\nonumber\\
&=&{r_+^2\over 4Gl^2T}-{M\over T}\nonumber\\ &=&-{M\over T}
+{J\Omega\over T}+{\pi r_+\over 2\hbar G}.
\end{eqnarray}
In the second line, the standard finite-temperature result that the range of
the $\tau$-integration is from zero to ${\hbar\over T}$ has been made use of.
This on-shell action is to be identified with the logarithm of the
grand canonical partition  function
and hence with $-{M-\Omega J-TS\over T}$. This
demonstrates the entropy at this level to be ${\pi r_+\over 2\hbar G}$.

Now the partition function is not really given by the exponential
of an on-shell action but involves  functional  integration  with
appropriate  boundary  conditions.  What  is  shown  above is the
leading term in the effective action. Quantum fluctuations about
the classical configuration lead to corrections.
A one-loop corrected action was given in \cite{CT} as
\begin{eqnarray}
-{I_E\over\hbar}=\int d\tau({r_+^2\over 4\hbar Gl^2}-{M\over\hbar})
+{2\pi r_+\over l}.
\end{eqnarray}
We rewrite this as
\begin{eqnarray}
-{I_E\over\hbar}&=&{r_+^2\over 4Gl^2 T}-{M\over  T}
+{2\pi r_+\over l}\nonumber\\ &=&
(1+{8\hbar G\over l}){\pi^2l^2T\over 2\hbar^2 G(1-\Omega^2l^2)},
\end{eqnarray}
where in the first line, we have again used the fact that the range of
the $\tau$-integration is from zero to ${\hbar\over T}$ and
in the last line we have rewritten the expression in terms of the
independent variables $T$ and $\Omega$ of the grand canonical
ensemble. It is assumed that the semiclassical values of these quantities
can continue to be used in this grand canonical description.
The above equation  implies that the one-loop corrected thermodynamic potential
is given by
\begin{eqnarray}
F=-(1+{8\hbar G\over l}){\pi^2l^2T^2\over 2\hbar^2 G(1-\Omega^2l^2)}.
\label{F}
\end{eqnarray}
Consequently,  the  entropy,  which  is  the  negative   of   the
temperature derivative of the thermodynamic potential at constant
chemical potential $\Omega$ is
\begin{eqnarray}
S&=&(1+{8\hbar G\over l}){\pi^2l^2T\over \hbar^2 G(1-\Omega^2l^2)}
\nonumber\\
&=&({\pi r_+\over 2\hbar G})(1+{8\hbar G\over l}).
\end{eqnarray}
This is the corrected entropy, and it is different from what one
would  na\"{\i}vely expect from the action because the correction
term is {\it twice} the change in the action.

The parameter $r_+$  which  occurs  in  this  expression  is  the
original  value  of $r_+$ and is related to the uncorrected mass
$M$ and the uncorrected angular momentum $J$ through (\ref{r}).
It has to be noted that $M,~J$ are no longer the correct physical
mass  or  angular  momentum  of the black hole. There are
corrected values which are easily calculated from (\ref{F})
using the formalism of the grand canonical ensemble:
\begin{eqnarray}
\tilde J=
-\left.{\partial F\over\partial \Omega}\right|_T=J(1+
{8\hbar G\over l}),
\end{eqnarray}
\begin{eqnarray}
\tilde M=F+TS+\Omega\tilde J=M(1+{8\hbar G\over l}).
\end{eqnarray}
It  is  now  possible to express the temperature and the chemical
potential in terms of these physical parameters. If a  corrected
$r_+$ is defined by
\begin{eqnarray}
\tilde r_+^2=4G\tilde Ml^2\left[1+\sqrt{1-
{\tilde J^2\over\tilde M^2l^2}}\right],
\end{eqnarray}
one has
\begin{eqnarray}
T={\hbar(\tilde r_+^2-4G\tilde Ml^2)\over \pi\tilde r_+l^2}
(1-{4\hbar G\over l}),
\end{eqnarray}
\begin{eqnarray}
\Omega={4G\tilde J\over\tilde r_+^2},
\end{eqnarray}
and the entropy can be rewritten as
\begin{eqnarray}
S=({\pi\tilde r_+\over 2\hbar G})(1+{4\hbar G\over l}).
\end{eqnarray}

\section{Discussion}
The corrections
that we are talking about involve extra powers of the Planck constant and are
therefore small compared to the semiclassical results for large systems. In the
present case, the condition for the corrections to be small is that
$l$ has to be large.
For large black holes, these $\hbar^0$
contributions are small relative to the area term where Planck's constant
appears in the denominator. On the other hand, for some extremal black hole
solutions obtained in string theory \cite{sen} the area of the
horizon vanishes. Quantum corrections will be especially important in these
cases.

To  sum  up,  we  have  demonstrated that the entropy of a black hole is
not simply given by the effective action.
However, if corrections
to the functional integral are known,  it is possible  to calculate
corrections  to  the  entropy  as well as to other thermodynamic
variables.

\end{document}